\input harvmac
\input epsf

\def\l{\lambda}
\noblackbox
\Title{\vbox{
\hbox{HUTP-98/A078}
\hbox{\tt hep-th/9811131}
}}{On the Gauge Theory/Geometry Correspondence}
\bigskip
\centerline{Rajesh Gopakumar \foot{gopakumr@tomonaga.harvard.edu}
and  Cumrun Vafa \foot{vafa@string.harvard.edu}}
\smallskip
\centerline{Lyman Laboratory of Physics}
\centerline{Harvard University}
\centerline{Cambridge, MA 02138}

\vskip .3in

The 't Hooft expansion of $SU(N)$
Chern-Simons theory on $S^3$ is proposed to be exactly
dual to the topological closed string theory on
the $S^2$ blow up of the conifold geometry.
The $B$-field on the $S^2$ has magnitude $Ng_s=\l$, the 't Hooft coupling.
We are able to make a number of checks, such as finding exact agreement at
the level of the partition
function computed on {\it both} sides for arbitrary $\l$ and to all orders in
$1/N$.  Moreover, it seems possible to derive this correspondence
from a linear sigma model description of the conifold.
We propose a picture whereby a perturbative
D-brane description, in terms of holes in
the closed string worldsheet,
arises automatically
from the coexistence of two phases in the underlying $U(1)$
gauge theory. This approach holds promise for a derivation of the
AdS/CFT correspondence.

\Date{November 1998}

\newsec{Introduction}

\lref\gopvaf{R. Gopakumar, C. Vafa, ``Branes and Fundamental Groups'',
hep-th/9712048.}
\lref\wittencs{E. Witten, ``Chern-Simons Gauge Theory As A String Theory'',
hep-th/9207094.}
\lref\witten{E. Witten, ``Quantum Field theory and the Jones Polynomial'',
Commun.Math.Phys. {\bf 121},(1989) 351 .}
\lref\doug{M. R. Douglas, ``Chern-Simons-Witten Theory as a
Topological Fermi Liquid '',
 hep-th/9403119}
\lref\twoacs{R. Gopakumar, C. Vafa, ``M-Theory and Topological Strings
 -I'',  hep-th/9809187}
\lref\witsuss{L. Susskind, E. Witten, ``The Holographic Bound in Anti-de
Sitter Space'', hep-th/9805114.}
\lref\thf{G. 't Hooft, ``On the Convergence of Planar Diagram Expansions'',
 Commun.Math.Phys.{\bf 86},(1982), 449.}
\lref\wittop{E. Witten,
``Topological Sigma Model,'' Comm. Math. Phys. {\bf 118} (1988) 411.}
\lref\momar{M. Marino and G. Moore, ``Counting higher genus curves in a
Calabi-Yau
manifold,'' hep-th/9808131.}
\lref\fabpand{C. Faber and R. Pandharipande, ``Hodge integrals and
Gromov-Witten theory,'' math.AG/9810173.}
\lref\koshen{K.Skenderis, M. Henningsson, ``The Holographic Weyl anomaly'',
hep-th/9806087.}
\lref\thoof{G. `t Hooft, ``A Planar Diagram Theory for Strong Interactions,''
Nucl. Phys. {\bf 72} (1974) 461.}
\lref\wittenl{E. Witten, ``Phases of N=2 Theories in 2 Dimensions,''
Nucl. Phys. {\bf B403} (1993) 159.}
\lref\bcovi{M.Bershadsky, S.Cecotti, H.Ooguri, C.Vafa , ``Holomorphic
Anomalies in Topological Field Theories, Nucl. Phys. {\bf B405} (1993), 279 .}
\lref\bcovii{M. Bershadsky, S. Cecotti,H. Ooguri, C. Vafa ,
``Kodaira-Spencer Theory of Gravity and Exact Results for
Quantum String Amplitudes'', Comm. Math. Phys. {\bf 165} (1994), 311 .}
\lref\peri{V. Periwal, `` Topological Closed-string Interpretation of
Chern-Simons Theory'', Phys. Rev. Lett. {\bf 71} (1993), 1295.}

The large $N$ limit of gauge theories has been much
studied since
`t Hooft \thoof ,
in the
hope of a string description.
The essential idea is the following: when the 't Hooft coupling
$\l=g^2_{YM}N <<1$ we have an ordinary Feynman diagram description
(with each diagram viewed as a Riemann surface with many holes).
The proposal has been that, in the
strong coupling limit $\lambda >>1$, there exists a closed string description
(the holes being somehow filled in or condensed).
A concrete and well studied example of this is large $N$ 2d Yang-Mills
\ref\gross{D. Gross, Nucl. Phys. {\bf B400}(1993),161; D. Gross, W. Taylor
Nucl. Phys. {\bf B400}(1993),181, {\bf B403}(1993),395}\
where a quasi-topological closed string description
\ref\moor{S. Cordes, G. Moore, S. Ramgoolam, ``Large N 2D Yang-Mills Theory
and Topological String Theory'', Commun. Math. Phys. {\bf 185}, (1997),
543; P. Horava, ``Topological Rigid String Theory and Two Dimensional
QCD'',
Nucl. Phys. {\bf B463}, (1996), 238;''On QCD String Theory and AdS
Dynamics'',
hep-th/9811028 .}\ emerges at large $N$  .
Some tentative attempts were also made in a similar direction
for large $N$ Chern-Simons theory \peri \doug .
Recently, the large $N$ limits of certain non-trivial
superconformal
gauge theories have been concretely proposed to be
closed superstrings on $AdS\times M$ backgrounds
\ref\mald{J. Maldacena, ``The Large N Limit of Superconformal Field Theories
and Supergravity,'' Adv. Theor. Math. Phys. {\bf 2} (1998) 231.}\ref\polgk{
S.S. Gubser, I.R. Klebanov and A.M. Polyakov, ``Gauge Theory Correlators from
Non-Critical String Theory,'' Phys. Lett. {\bf B428} (1998) 105.}\ref\witte{E.
Witten, ``Anti De Sitter Space And Holography,'' Adv. Theor. Math. Phys. {\bf
2} (1998) 253.}.
Extensions of the proposal in various directions as well as
many non-trivial tests have been successfully carried out.

The aim of this paper is twofold: First, we propose a new duality --
the large $N$ limit of $SU(N)$ Chern-Simons theory on $S^3$ is {\it exactly}
the
same as an $N=2$
topological closed string on the $S^2$ blow up of the conifold geometry.
Since both sides of the duality have been relatively well studied, we can
compare, for instance, the partition functions on both sides.
We find a strikingly exact match for all $\l$
(which plays the
role of $1/\alpha^{\prime}$) and to all
orders in $1/N$. (In \twoacs , we had computed the closed string
partition function for the $S^2$
blowup geometry of the conifold using M-theory, and had partially anticipated
the connection to Chern-Simons theory which we fully develop here.)
We are also able to successfully compare the
coupling of the gauge theory to gravity, with
quantities on the closed string side. This will also exhibit the by now
familiar $UV/IR$ relation \witsuss\ .
Though we have not made a detailed
comparison, observables such as Wilson loops in the gauge theory (which
compute knot invariants) should be given by holomorphic surfaces
in the $S^2$ blown up geometry with a boundary
which approaches a
 knot configuration on the $S^3$ at infinity.

This conjecture is very much in the
spirit of the AdS/CFT correspondence
in that the $S^3$ Chern-Simons theory
has
a topological {\it open} string description where we put $N$
(topological) 3-branes on an $S^3$
inside a conifold $T^* S^3$ background
\wittencs .
And the dual description we propose, is in terms of the same topological
string,
but now with no open strings,
and on a modified background with parameters depending
on $\lambda =g_sN$.  In our case, the complexified area $t$ of the $S^2$
(i.e. with the
B-field as its imaginary part)
has the identification $t=ig_sN=i\lambda$. Intuitively, one can
think of the $S^2$ as arising from the cotangent direction of $T^*S^3$
and corresponds to the two-sphere surrounding the positions of the 3-branes
in this transverse ``$R^3$''.
One may also arrive at this correspondence, as we will explain in section
3, by considering the coupling of the Chern-Simons theory
to gravity in the bulk.  This will be seen to be quite similar
to the way in which the conformal anomaly of the $N=4$ Super Yang-Mills
manifests itself on the gravity side.

However, here we can actually do more:
We suggest a simple {\it derivation}  of this
correspondence by considering a 2d QFT on
the closed string side, which flows in the IR to the conformal field
theory on the worldsheet. Examining this QFT
in the limit $g_sN\propto t\rightarrow 0$, we find that there are two
coexisting phases.
In one phase (``Coulomb'') the theory
is free abelian in the IR and can be integrated out, leaving
``holes'' on the worldsheet. While in the other (``Higgs''), it is non-trivial
and gives the bulk of the worldsheet theory.  Moreover the worldsheet
fields
in the Higgs phase, satisfy Dirichlet conditions at the boundaries,
which gives rise to an equivalent D-brane description in this limit.

This story parallels a similar one
in the study of superstring vacua:  The
correspondence (duality) between Landau-Ginzburg/Calabi-Yau
sigma models \ref\gvw{B. Greene, C. Vafa and N. Warner,
``Calabi-Yau Manifolds and Renormalization Group Flows,''
Nucl. Phys. {\bf B324} (1989) 371.}\ref\mart{E. Martinec,
``Criticality, Catastrophes, and Compactifications,'' in
{\it Physics and Mathematics of Strings}, ed. L. Brink,
D. Friedan and A.M. Polyakov (World Scientific, 1990).}\ was explained in
\wittenl\ by
studying a one parameter family of 2d QFTs with extra degrees of freedom in
the UV.
The resulting  CFT's in the IR provide a smooth interpolation between the
two descriptions.
In fact, the 2d QFT (linear sigma model)
description of the $S^2$ resolution of the conifold, which we
use in this paper, was already
proposed in \wittenl .
This linear sigma model is just a $U(1)$ gauge theory
with a single phase, the Higgs phase, when  $t \not =0$.
As $t \rightarrow 0$
the Coulomb branch also opens up.
Moreover, the topological string amplitudes,
which are dominated by both non-linear sigma model instantons and $U(1)$
gauge instantons (vortex configurations), consist in this limit of only the
latter.
These vortices, which,
in the generic Higgs phase, are point-like in the IR limit, can, in the
$\lambda =0$ limit, have an arbitrary size. It is the ``core''
of these vortex lines that we will identify with the
filled holes of worldsheet D-branes. We believe this procedure of replacing
D-brane rules of  worldsheet amplitudes in terms of a closed string theory
description is bound to have other
applications.\foot{The idea of condensing Dirichlet
holes using extra degrees of freedom
on the worldsheet, was already anticipated in \ref\Grepo{M. Green and J.
Polchinski, ``Summing Over World-Sheet Boundaries,''
Phys. Lett. {\bf B335} (1994) 377.}\
(as was pointed out to us by Albion Lawrence). Even though the precise
description we find differs from theirs,
(we have a {\it two phase system} which generates the surfaces with holes),
there are some resemblances between the two approaches.}
In particular, it might be possible
to prove the AdS/CFT conjectures along such lines.

The organization of this paper is as follows:  In section 2 we state
the large $N$ conjecture for $SU(N)$ Chern-Simons theory on $S^3$.
In section 3 we perform the above mentioned explicit
checks on this conjecture as well
as provide further rationales for it.  In section 4 we outline a derivation
of this correspondence in terms of the 2-d linear sigma model.
Finally in section 5 we make suggestions for some generalizations
of the ideas presented here.

\newsec{The Conjecture}

Before going into the details of the particular
conjecture which is the subject of this paper,
it will be useful to set out some generalities regarding the open string
description of large $N$ gauge
theories and their conjectured  relation to a closed string dual.
This will also help in highlighting the similarities between our
present conjecture and the AdS/CFT ones.

Take any large $N$ gauge theory containing only fields in the adjoint.
The perturbative Feynman diagrams that
contribute to the free energy (or more generally,
gauge invariant correlation functions)
admits the following 't Hooft organization:
\eqn\tHooft{F=\sum_{g=0,h=1}C_{g,h}N^{h}\kappa^{2g-2+h}
=\sum_{g=0,h=1}C_{g,h}N^{2-2g}\lambda^{2g-2+h}}
Here  $\kappa$ is the ordinary gauge coupling ($g^2_{YM}$ for a Yang-Mills
theory and ${2\pi\over k+N}$ in the Chern-Simons theory), while
$\lambda=\kappa N$ is the 't Hooft coupling which is held
fixed in the large $N$ limit. In writing eqn. \tHooft , we have used the
fact that a Feynman diagram in the double line notation can be thought of
as a triangulation of a closed Riemann surface. Thus, in a diagram with $h$
faces,
$V$ vertices and $E$ edges, we get the factor of $N^h\kappa^{E-V}$.

This also looks like an open string expansion on worldsheets with $g$
handles and $h$ boundaries. The double line notation is here seen as
representing the world lines of the endpoints of the open string.
In fact, in the case of Chern-Simons theory on a 3-dimensional
manifold $M$, \foot{ Chern-Simons theory on $M$ is defined via the
action  $S_{CS}= {k\over 4\pi}\int_M{\rm Tr}(AdA+{2\over 3}A^3)$. See
\witten\ for further details.}
it was shown by Witten \wittencs\ that the coefficient
$C_{g,h}=Z_{g,h}$, where $Z_{g,h}$ is the partition function of an open
string theory on a worldsheet with $g$ handles and $h$ boundaries.
The theory in this case is the A-model topological open string theory
with $N$ ``topological'' D-branes on $M$ in an ambient
six dimensional target space $T^*M$.

A similar thing happens in $N=4$ Super Yang-Mills in $d=4$. (Though
the free energy vanishes here due to supersymmetry, correlation
functions have a similar expansion. The $C_{g,h}$'s will now carry
momentum dependence as well.) This theory arises as the low energy
limit of Type IIB open strings in flat space with a background of $N$ D-3
branes.
And we can once again identify the genus $g$ and holes $h$ of the gauge
theory diagram with those of the worldsheet.

In all these cases, one might ask as to what happens when one carries out the
sum over the holes $h$ in eqn.\tHooft\ (or the appropriate generalization).
We expect something of the form
\eqn\thft{F=\sum_{g=0}N^{2-2g}F_g(\l),}
which is more like a closed string expansion. The question, since 't Hooft,
has been: what is this closed string theory?\foot{In
carrying out this sum we have assumed the existence of a radius of
convergence. This is by no means obvious. 't Hooft has established nice
convergence properties for planar diagrams in a number of field theories
\thf . And we will explicitly see this for all genus in the Chern-Simons
theory. But even with a finite radius of convergence we could
run up against the issue
of a phase transition separating the perturbative and the stringy regime
\ref\peo{M. Douglas, V. Kazakov, `` Large $N$ Phase Transition in
Continuum QCD$_2$'', Phys. Lett. {\bf B319}, (1993), 219.}.}
Maldacena has proposed an answer in the case of $N=4$ $SU(N)$ Super
Yang-Mills.
It is closed
IIB string theory on a background which is $AdS_5\times S^5$ with an RR field
and a curvature scale set by $\lambda$ and with $g_s\propto{\l\over N}$.
Despite all
the
remarkable amount of evidence for it, we do not yet have a direct demonstration
of this statement. We note the striking feature of this conjecture that by the
time we have summed over all the holes the background has transmuted
from $D3$ branes in flat $R^{10}$, to the curved $AdS_5\times S^5$
without any branes.  In both regimes, there exists a string theory
description and this conjecture can be viewed as a statement about
a 2d QFT on the worldsheet which interpolates between them.

Here, we will propose an answer for the case of
$SU(N)$ Chern-Simons theory on $S^3$. The claim is that it is exactly dual to
the A-model topological closed
string theory on the $S^2$ resolved conifold geometry. Let us first briefly
review what this means.

The A-model topological closed string theory on a Calabi-Yau manifold $M$
is described by a ``twisted'' sigma model on $M$ \wittop . This enormously
simplifies the theory and the contribution to the string path integral
comes solely from holomorphic maps $X(z)$ onto $M$. Besides the constant
maps, these consist of `instantons' which are topologically nontrivial
mappings
from the genus $g$ worldsheet onto two dimensional surfaces in $M$.
We will write down the
resulting structure of the partition function for genus $g$ in the
next section, when we make a detailed comparison. The other feature of the
topological A-model that we will need is that it is independent of the
complex structure deformations of $M$ and depends only on the kahler ones.
The particular Calabi-Yau $M$ that we are proposing in our dual is the
local geometry near a conifold singularity which has been resolved
by an $S^2$. We will be led to the identification
of $\l$ with the B-field flux through the $S^2$ and $g_s={i\l\over N}$.

As we mentioned earlier, the $SU(N)$ Chern-Simons theory itself arises
from the open string version of the topological A-model in the presence of
D-branes. In fact, the A-model open string theory on the Calabi-Yau
$T^*S^3$ with Dirichlet boundary conditions on the $S^3$ gives rise to
the $SU(N)$ gauge theory on $S^3$ in the manner outlined below Eq.\tHooft .
 The $T^*S^3$ geometry happens to be the other side of the conifold. In
other words when the singularity has been resolved by an $S^3$. So
we
see that, just as in the AdS/CFT cases,
summing over holes has made the original $T^*S^3$ undergo the
conifold transition to the resolved geometry with no branes. (See Fig.1)

\bigskip
\centerline{\epsfxsize 6.truein \epsfysize 1.5truein\epsfbox{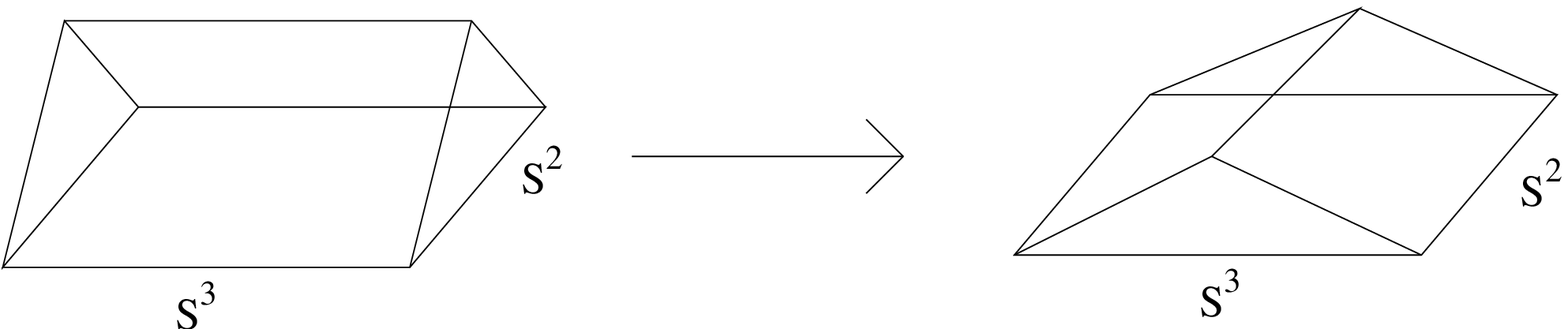}}
\noindent{\ninepoint\sl \baselineskip=8pt {\bf Fig.1}: {\rm
The geometry of $T^*S^3$ with an $S^3$ of finite size goes into an $S^2$
resolved geometry after the conifold transition. }}

As the figure indicates, the geometry of the conifold is essentially
like a cone with a base which is topologically $S^2\times S^3$.
In the $S^2$ resolved  geometry on the right, the space at infinity
is $S^2\times
S^3$ with the $S^2$ of finite size.
This is like the analogous $S^5$ of the Maldacena conjecture.
The gauge theory itself can be thought of as living on the
large $S^3$.  In fact one can push the analogy even further:
In the $AdS^5\times S^5$ description of $N=4$ Yang-Mills, (in Euclidean
version)
the boundary is $S^4\times S^5$ with the radial direction of $AdS_5$
filling in the $S^4$.  In the weak coupling regime, the boundary
is still $S^4\times S^5$ but the difference is that now it is the
$S^5$ that gets filled.  So in some sense there is already
a conifold like transition in the $N=4$ Yang-Mills as well, when
we go from weak to strong coupling.
In fact, Fig.1 for the conifold is topologically
 also accurate for this
case if we replace $S^3\rightarrow S^4$ and $S^2\rightarrow S^5$.

More detailed aspects of the conifold
geometry and its blow up will be discussed in section 4.  Here
we would like to remark that, in a certain geometrical sense
we are exactly at the transition point. On the gauge theory
side the amplitudes do not depend on the size of the $S^3$ (as it is
topological) and so
we can take it to have shrunk to zero size. On the closed string
side, the actual area of the $S^2$ is zero --  it is only the $B$-field
which is turned on.  This will prove important in deriving
the D-brane rules from the closed string theory description,
as we will see in section 4.

Let us remark in passing that this conjecture provides an example
of a closed string dual of a non-supersymmetric gauge
theory that lives in {\it three} (rather than one) higher dimensions.

\subsec{Wilson Loop Observables}

Next we turn our attention to Wilson loops in the Chern-Simons theory
which are the celebrated knot polynomials \witten .
Consider a loop $\Gamma$ in $S^3$ and the Wilson loop
observable of the Chern-Simons theory
$<W(\Gamma )>=<{\rm Tr} P{\rm exp}(i\int_\Gamma A )>$
in the fundamental representation of $SU(N)$. We can generalize this
by considering arbitrary representations as well as an arbitrary number
of Wilson loops.  For simplicity let us restrict our attention
to a single loop in the fundamental representation.
We would
like to give a prescription for computing it on the closed string side.
In order to do this,
we will need some facts about the topological A-model on worldsheets with
boundary, which we will now review.

 It is known that the boundary should be mapped onto
a Lagrangian 3-dimensional submanifold $C_3\subset M$ of the
Calabi-Yau 3-fold \wittencs .  Being Lagrangian means
that the Kahler form $k$ vanishes when restricted to $C_3$.
As we said earlier, the path-integral in the A-model is dominated by
holomorphic
Riemann surfaces. Then the condition that these end on a Lagrangian
submanifold implies that the neighborhood near the boundary is mapped
to a half cylinder which is orthogonal to $C_3$.

Let $f$ be an isolated
holomorphic map from the worldsheet $\Sigma$ to the Calabi-Yau $M$
with one boundary on $C_3$. The fact that it is holomorphic implies
that the area is given by
$$A(\Sigma )=\int_\Sigma f^*(k)$$
where $f^*(k)$ denotes the pull back of $k$ on the worldsheet.
Such a map will have a contribution of
$\pm {\rm exp}(-A)$ to the partition function (the $\pm$ is determined
by certain ratio of determinants \wittencs ).  Now,
given another holomorphic map $f'$, with a similar
$\pm {\rm exp}(-A')$ contribution, let us look at the difference $A'- A$.
Let $\gamma$ and $\gamma'$ denote
the boundaries of $\Sigma $ and $\Sigma '$ in $C_3$
respectively.  Consider a 2-dimensional submanifold
$B_2\subset C_3$, whose boundary is $\partial B_2=\gamma -\gamma'$
and the closed 2-surface ${\hat \Sigma}$, in $M$; ${\hat \Sigma}\equiv
f(\Sigma) +B_2+f(\Sigma ')$.  It is easy to see that
with opposite orientations on $f(\Sigma)$ and $f(\Sigma ')$,
the boundary of ${\hat \Sigma}$ is empty.
Let us consider the integral of $k$ on $\hat \Sigma$:
$$\int_{\hat \Sigma}k=-A +\int_{B_2}k+A'=A'-A$$
where we have used the fact that $B_2\in C_3$ and that $k$ vanishes
on $C_3$. Thus, we learn that the difference $A'- A$
is given by the integral of $k$ on a closed 2-cycle $\hat \Sigma$
in $M$, which is thus quantized. In particular, if we have only one Kahler
class, parameterized by the Kahler parameter $t$, then
$$A'-A=n t$$
for some integer $n$.

With these preliminaries we are now ready to define a prescription
for the computation of
$W(\Gamma )$ in the $S^2$ blown
up geometry.  As in \ref\juanwil{J. Maldacena, `Wilson loops in large
N field theories'',Phys.Rev.Lett. {\bf 80}, (1998), 4859; S-J. Rey, J.Yee,
`` Macroscopic Strings as Heavy Quarks of Large N Gauge Theory and
Anti-de Sitter Supergravity'', hep-th/9803001}\
it is natural to consider worldsheets
whose boundary is a loop $\Gamma$ living on the boundary of the space.
In our case the boundary is topologically
$S^2\times S^3$.  Fix a large $S^3$ of size
$\Lambda$ on the boundary,
which we take to be a Lagrangian submanifold.  We consider holomorphic
maps from the worldsheet with boundary and require the boundary
to lie in $S^3_\Lambda$, and that the boundary intersect
$S^3_\Lambda$ on the knot $\Gamma$ (or on a knot isotopic
to $\Gamma$)\foot{Actually, the condition that the boundary be mapped to a
fixed curve $\Gamma$
is too strong \ref\taubes{C. Taubes, private communication}\ and
no such holomorphic map would generically exist. It
seems more natural to fix the knot type $\Gamma$ in $S^3_\Lambda$. And then
require that the boundary of the worldsheet
be any loop $\Gamma'$ which
is smoothly deformable to $\Gamma$ (i.e. $\Gamma'$
and $\Gamma$ are equivalent knots)}.  This is our proposal for computing
Wilson loop observables on the closed string side.

Note that the area $A$ of surfaces ending on $S^3_\Lambda$
will depend on $\Lambda$.  Moreover, as $\Lambda \rightarrow \infty$
this area will diverge.  However the {\it difference} between
two surfaces ending on $S^3$ will be independent of $\Lambda$ and finite,
as explained above. Thus the proposed topological A-model observables
are well defined up to an overall multiplication by
$exp(-A_0(\Lambda))$. This is in fact to be expected, since
on the Chern-Simons side, the Wilson loop observables
are divergent due to UV singularities. They can be regularized
by a choice of point splitting, known as `framing' the knot \witten .
Different framings alter $W(\Gamma)$ by multiplicative
factors which we are now identifying with the choice in the overall
area subtraction. In which case, remembering the quantization of area
differences, we have the proposal
\eqn\dwil{<W(\Gamma )> ={\rm exp }(-A_0)
\sum_{\del f_n(\Sigma_g)\sim \Gamma}\pm (g_s)^{2g-1}exp (-nt)}
where $f_n$ are holomorphic maps from a surface $\Sigma_g$ of genus
$g$ with one boundary on $\Gamma$.
If the holomorphic curves come in families one is expected to compute
certain integrals over the moduli space of such curves, and the sum above
should be understood in that sense. In the next section we will see some
evidence that knot invariants on $S^3$ do indeed have this structure.

\newsec{Checks of the Conjecture}

The most striking check of this conjecture is the matching of the free
energies. In other words, the genus $g$ contribution in \thft\ of the
Chern-Simons theory,
will precisely match the genus $g$ partition function
of the closed string theory on the $S^2$ resolved geometry. Next, we
consider the anomalous
gravitational coupling of the Chern-Simons theory and relate
it to a similar ``anomaly'' on the closed string side.
Finally, we also consider Wilson loop
observables on the gravity side.  In this case,
since the configuration of holomorphic maps of interest
have not been previously
studied, we cannot check the statements in detail. We will, however,
demonstrate that the structure of the knot invariants agrees with the
general structure expected from the computations on the closed string
side.  Moreover, having gained confidence in our conjecture,
we can instead turn this around and propose a reformulation of knot
invariants in terms of a summation over minimal
surfaces bounding the knot.

\subsec{The Free Energy}

The exact
free energy of the $SU(N)$ Chern-Simons theory on $S^3$
\peri\ can be expanded in
an
't Hooft expansion in a fairly straightforward way. The main point to be
remembered is that the bare 't Hooft coupling $\lambda_b = {2\pi N\over
k}$
is finitely renormalized to $\lambda={2\pi N\over k+N}$ (as follows from
the renormalization of $k$ \witten ).
The free energy contribution to $N^{2-2g}$, written in the open string
expansion
of Eq.\tHooft\ , is (See appendix of \twoacs\ for details. Her we will
define our free energy via $Z=e^{-F}$, differing in overall sign from
\twoacs )
\eqn\Fg{F_g(\l)=-{\chi_g[1+2\sum_{p=1}^{\infty}\zeta(2g-2+2p)
{\eqalign{\pmatrix{2g-3+2p\cr 2p}}}
({\l\over 2\pi})^{2g-2+2p}]}}
for $g>1$ \foot{Note that the exact Chern-Simons
expression has a $\l$ independent
term. This and the $\ln\l$ terms in $F_{0,1}$ are present in the appendix
of \twoacs\
but were not interpreted there. }.
Therefore the  coefficients $C_{g,h}=Z_{g,h}$ in \tHooft\
have the simple form
\eqn\cgh{C_{g,2p}=-\chi_{g,2p}{2\zeta(2g-2+2p)\over (2\pi)^{2g-2+2p}}
;\quad C_{g,2p+1}=0.}
Here
\eqn\euler{\chi_{g,h}= (-1)^h{\eqalign{\pmatrix{2g-3+h\cr h}}}\chi_g =
(-1)^h{\eqalign{\pmatrix{2g-3+h\cr h}}}(-1)^{g-1} {B_g\over 2g(2g-2)}}
is the Euler characteristic of the moduli space of Riemann surfaces with
genus $g$ and $h$ punctures.

The similar answer for genus zero and one (the coefficients of $N^2$ and
$N^0$ respectively) takes the form
\eqn\zero{F_0(\l)= {3\over 4}-{1\over 2}\ln\l-\sum_{p=2}^{\infty}
{\zeta(2p-2)\over p-1}{({\l\over
2\pi})^{2p-2}\over 2p(2p-1)}}
\eqn\one{F_1(\l)= -\sum_{p=1}^{\infty}B_1{\zeta(2p)\over
2p}({\l\over 2\pi})^{2p}.}

The sum over the number of holes $h=2p$ can be carried out in all these
cases.

{\bf Genus zero:} The sum in Eq.\zero\ can be carried out
(after taking two derivatives and rewriting $\zeta(2m)=\sum_{n=1}{1\over
n^{2m}}$.)
\eqn\zerosum{N^2F_0(\l)=-({N\over \l})^2[-\zeta(3) +i{\pi^2\over 6}\l
-i(m+{1\over 4})\pi\l^2 +{i\l^3\over 12}
+\sum_{n=1}^\infty {e^{-in\l}\over n^3}]}
(the integer $m$ in the above expression
is not uniquely fixed--this is also echoed
on the closed string side as we will note below).
With the identifications, $g_s={i\l\over N}$ (the $i$ comes from the $i$ in
the Chern-Simons action) and $t=i\l$ for the
complexified Kahler parameter for the $S^2$, our answer reads as
\eqn\zerotop{{\cal F}_0={1\over g_s^2}[-\zeta(3) +{\pi^2\over 6}t
+i(m+{1\over 4})\pi t^2 -{t^3\over 12}
+\sum_n {e^{-nt}\over n^3}].}
Let us compare this with a general genus 0 topological
string amplitude for a Calabi-Yau with one Kahler
class \ref\cand{
P. Candelas, X.C. de la Ossa, P.S. Green and L. Parkes, ``A
Pair of Calabi-Yau Manifolds as an Exactly Solube Superconformal
Theory,'' Nucl. Phys. {\bf B 359} (1991) 21.}
\eqn\fzero{{\cal F}_0={1\over g_s^2}\big [-{\chi \over 2}\zeta(3)-
{\pi^2 \over 6}c_2t + i\pi at^2 -
C {t^3\over 3!}+\sum_{n,m}d_{m}{1\over n^3}{\rm exp}(-nmt)\big ].}
Here  $\chi$ is the Euler
characteristic of the manifold\foot{The coefficient $\zeta(3){\chi\over 2}$ has
its origin
in the $R^4$ term of type II strings.},
$c_2$ the second Chern class
(more precisely, it is  $c_2t=\int k\wedge c_2 $ that appears here
\ref\yauet{S. Hosono et. al., ``Mirror Symmetry, Mirror Map
and Applications to Complete Intersection Calabi-Yau Spaces,''
Mirror Symmetry II, ed. by B. Greene and S.-T. Yau,
AMS/IP.}) and $C$ is
the classical
self-intersection number for the Kahler class (i.e. $Ct^3=\int_M k\wedge k
\wedge
k$). Also $d_m$ denotes the number
of primitive degree $m$ holomorphic spheres, and $n$ labels the
multi-covering of the basic ones.
The quantity $a$ in the above formula does not
have a known topological interpretation;
However for the case of Calabi-Yau with
one modulus, it is predicted \yauet\ that $a=C/2$ mod ${\bf Z}$.\foot{From the
viewpoint of
closed string theory, only the third derivative and higher of ${\cal F}_0$
(and first and higher of ${\cal F}_1$) refer to physical
quantities. However, mirror symmetry naturally dictates the above form
\yauet\ .}

The various terms in our expression
\zerotop\ are now easily recognized!
Let us first recall that in the $S^2$ resolved geometry
there is only one primitive
holomorphic sphere and it is of degree one, so that $d_1=1$ and $d_i=0$
for $i>0$.  Moreover, we are led to the identifications:
$$C={1\over 2}, \qquad c_2= -1,\qquad \chi =2.$$
To compare these quantities with those in our $S^2$
blown up geometry, we have to recall that, for non-compact manifolds,
some of
these quantities are naively divergent and have to be carefully regularized.
(This point will be crucial to the discussion in the next subsection).
But we can nevertheless see why these assignments are natural for our geometry.
The formal continuation of  $t\rightarrow -t$ corresponds to a ``flop''
of the blow up geometry, under which the
self-intersection number, of the $S^2$,
undergoes a down shift by 1 \wittenl\ref\asmg{P. Aspinwall, B. Greene
and D. Morrison, ``Calabi-Yau Moduli Space, Mirror Manifolds
and Spacetime Topology Change in String Theory,'' Nucl. Phys.
{\bf B416} (1994) 414.}. By requiring that the above expansion be
valid on both sides of the flop, we fix $C=1/2$
(note that $t^3$ is odd under $t\rightarrow
-t$).  A similar  argument shows \ref\katz{S. Katz, private
communication.}\ that under the flop $\int c_2\wedge k$
goes up by $2t$ which fixes $\int k\wedge c_2=-t$
as a valid formula on both sides of the flop.
 The value of $\chi=2$ is also natural: $\chi$
is twice the difference between the
number of Kahler and complex
deformations.  In the case at hand, we have only one of the former and none
of the latter, giving  $\chi =2$.  Note that $\chi$ does not
change under a flop, as expected.

{\bf Genus one:}
The genus one answer, \one\ can also be easily written in closed form as
\eqn\onesum{N^0F_1(\l)= i{B_1\over 4}\l
+{B_1\over 2}\ln(1-e^{-i\l})}
Once again, with the above identifications we get
\eqn\onetop{{\cal F}_1 = {1\over 24}t
+{1\over 12}\ln(1-e^{-t}).}
Again, we should compare this with the genus one
answer of \bcovi\ (specialized to the case that there are no $g>0$
holomorphic curves)
\eqn\gbc{{\cal F}_1=-{c_2\over 24}t+{1\over 12}\sum_m d_m\ln  (1-e^{-mt})}
In our case, with the $d_i=\delta_{i1}$ and $c_2=-1$ as above, \gbc\ reduces
precisely to \onetop .

{\bf Genus $g>1$ :}
And finally, the non-trivial structure of higher genus terms
(the sum over $p$ in \Fg\ is carried out in \twoacs ) is captured in
\eqn\Fgsum{N^{2-2g}F_g(\l)=({N\over \l})^{2-2g}(-1)^{g-1}[(-1)^{g-1}\chi_g
{2\zeta(2g-2)\over (2\pi)^{2g-2}}-
\chi_g\sum_{n \in {\bf Z}}{1\over (\l+2\pi n)^{2g-2}}]}
which after rewriting the sum and translating into closed string variables
reads as
\eqn\Fgtop{{\cal F}_g= g_s^{2g-2}[(-1)^{g-1}\chi_g
{2\zeta(2g-2)\over (2\pi)^{2g-2}}-
{\chi_g\over(2g-3)!}\sum_{n=1}^{\infty} n^{2g-3}e^{-nt}].}
The structure of the closed topological string anticipated
 in this case is \bcovii :
\eqn\gstru{{\cal F}_g= g_s^{2g-2}[{\chi \over 2}\int_{{\cal M}_g}
c_{g-1}^3+\sum \alpha_n e^{-nt} ]}
  The first term
in brackets
is the contribution from constant maps -- the whole worldsheet is mapped
to a point.  Here
$c_{g-1}^3$ is a certain characteristic class
on the  moduli space of genus $g$ Riemann surfaces.  The second
term corresponds to the contribution of genus $g$ curves covering the sphere
$n$ times and $\alpha_n$ are some universal
coefficients.   The contribution $\int_{{\cal M}_g} c_{g-1}^3$ from
constant maps
was recently obtained using
the M-theory/
type IIA correspondence
\twoacs\ and by using type IIA/heterotic duality \momar\
to be exactly given by
$$\int_{{\cal M}_g} c_{g-1}^3=(-1)^{g-1}\chi_g
{2\zeta(2g-2)\over (2\pi)^{2g-2}}.$$
With the $\chi =2$ in our case, we see that the first term in \gstru\
matches with the first term in \Fgtop .  This term has a simple
physical interpretation \twoacs . It is seen as coming from a one loop
integrating out ({\it a la} Schwinger) of
charged zero branes in the physical IIA theory on the
Calabi-Yau.  The integral $\int_{{\cal M}_g} c_{g-1}^3$
was also derived via a direct computation by mathematicians \fabpand .
The coefficients $\alpha_n$, corresponding
to the contribution from multiply wound higher
genus curves over $S^2$, were also derived using the Schwinger
one-loop
computation \twoacs\ . The final answer
exactly matches that predicted in \Fgtop (compare with Eqn. (3.3) in
\twoacs ).
In fact, the form of the second term presented in \Fgsum\
is that obtained by integrating out bound states of $S^2$
wrapped two branes and zero branes in the physical IIA theory.  The
$\alpha_n$ have again been computed directly by
mathematicians \fabpand\ and the result agrees with that obtained from
the Schwinger computation.
Thus once again the large $N$ free energy agrees fully with the topological
string partition function on the $S^2$ blow up!

\subsec{The Coupling to Gravity}

While the above detailed agreement of the free energies should convince even
hard-boiled skeptics, we can actually match a few more quantities.
In fact we can ask as to how the topological branes couple to gravity,
which is after all the reason why the geometry is altered.

Formally one expects  that
the Chern-Simons theory is independent of the background metric. Indeed,
the classical Lagrangian is background independent.
However, it was shown in \witten\ that there can be a coupling
at the quantum level via the gravitational Chern-Simons term. It
was argued that, in a particular regularization
one would need to add the gravitational
Chern-Simons term as a counterterm in order
for the Chern-Simons partition function
to be purely topological.  In other words, the free energy of
Chern-Simons theory $F_{CS}$ is
\eqn\gravcs{F_{CS}=F_{Top}+ i\pi {c\over 12}I(\omega)}
where
$$I(\omega)={1\over 8\pi^2}\int_{S^3}Tr(\omega d\omega
+{2\over 3}\omega^3).$$
is the gravitational Chern-Simons action in terms of $\omega$
which is the spin connection and
$c$ is the central charge of the $SU(N)$ WZW CFT at level $k$,
\eqn\cent{c={k(N^2-1)\over k+N}= -{1\over 2\pi}(N^2-1)(\l - 2\pi).}
This result was motivated in \witten\ from a one-loop
calculation.  It has also been confirmed at the two loop
level \ref\axsin{S. Axelrod and I. Singer, ``Chern--Simons Perturbation Theory
II,'' J. Diff. Geom. {\bf 39} (1994) 173.}.  Furthermore, it was conjectured
in \wittencs\ that this should be the only coupling
between the closed string sector of the topological A-model with the open
string sector that gives rise to the
Chern-Simons theory.

The divergences in perturbative gauge theory which give rise to the
gravitational coupling come from the ultraviolet.
In the spirit of AdS/CFT correspondence one expects
this to show up as an IR effect on the closed string side \witsuss .
Indeed, there are
such potential IR divergences. Consider, for example,
topological A-model at genus one.  It was shown in \bcovi\ that
the leading term in the large area limit involves a term
$${\cal F}_1= {1\over 24}\big[{1\over 8\pi^2}\int k \wedge {\rm tr}( R\wedge R
)\big ]+...$$
For compact manifolds this is a topological invariant
in the form of ${1\over 24}\int k\wedge c_2$
where $c_2$ is the second Chern class of the manifold. However,
for manifolds with boundary this is not a topological invariant.  This
is similar to the better known case of $\int R$ on 2-manifolds
with boundaries, where it is no longer a topological invariant
unless we add a boundary term.
In the case at hand, the boundary is
topologically $S^3\times S^2$ and it is the quantity
\eqn\ctwotop{\int k\wedge c_2\rightarrow
{1\over 8\pi^2}\int k \wedge {\rm tr} (R\wedge R) -\int_{S^3}
I(\omega)\int_{S^2} k }
which is topological, where the $S^2\times S^3$
is identified with the boundary of the blow up geometry
of the conifold.  To see the topological invariance of this
quantity, one uses the relation ${\rm Tr} (R \wedge R)= d
L_{CS}(\omega)$.  This implies that the topological closed string theory
also has an ``anomaly'' on non-compact manifolds -- one obtains a surface
term which depends on the boundary metric.
\eqn\gravtop{{\cal F}_1 ={\cal F}_{1}^{Top} +{1\over 24}I(\omega)\int_{S^2}
k .} %
This matches the structure we found in the Chern-Simons theory eqn \gravcs
.  Note that the dependence of the coefficient in \gravcs ,\cent\ predicts
a genus 0 (the $N^2$ term) as well as a genus 1 contribution
($N^0$ term).  Let us first consider the genus 1 contribution
and compare it with what we are finding from the topological
string theory at genus 1.
To match the coefficient, we first need to shift $\l
\rightarrow \l+2\pi$ in Eqn \cent\ .\foot{ This is the shift of the B-field
by $2\pi$. The instanton terms in brackets in \onesum\ and \zerosum\ are
obviously invariant under this shift. It is easy to check that the
topologically meaningful coefficients of the constant,
the linear and the
cubic terms in \zerosum\ are also left unchanged. The only changes
are in the anyway ambiguous quadratic and constant terms in ${\cal F}_0$ and
${\cal F}_1$ respectively.} We then see that the coefficient of $N^0I(\omega)$
is
$i{\l\over 24}$ matching that in \gravtop\ , with the identification
$t\equiv\int_{S^2} k=i\l$. Thus, the gravitational coupling has given us an
independent derivation of this identification, or depending on your view,
another check. Note that it is crucial that the group is $SU(N)$ rather
than $U(N)$ for this match to work.  This is also similar to
what has been found in the AdS/CFT correspondences \witte .

But we also have a genus zero contribution in Eqn.\cent\ , i.e.
\eqn\zerograv{\delta_{grav} F_0= -{i\over 24g_s^2}(\l+2\pi)^2\l I(\omega)}
where we have kept in mind the shift in $\l$ and that $({N\over
\l})^2={1\over g_s^2}$. We firstly note that there is no constant term,
signifying that we do not get an $I(\omega)$ contribution from regularizing
the Euler character term in \fzero . But we do have a linear term like in
genus one which comes from the corresponding linear term in \fzero . Noting the
substitution in \ctwotop , we see that
the closed string again has the anomalous
contribution
$$\delta{\cal F}_0 = -{1\over g_s^2}{\pi^2\over 6}t I(\omega).$$
This is again exactly what we see on the Chern-Simons theory side in the
linear term of Eqn. \zerograv .
The quadratic term was anyhow ambiguous and did not seem to have a
topological meaning. The cubic term in \zerograv\ is presumably seen on the
closed string side after regularizing the topological $C=\int k\wedge
k\wedge k$ term.  It would be interesting to verify this.

To summarize, we find that the gravity counterterms in the gauge
theory are reflected in similar non-topological infrared contributions
on the closed string side. The anomalous coupling to gravity in
Chern-Simons theory
may be thought of as a simple analogue of the conformal anomaly in a conformal
field theory which, again, arises from the regularization of UV divergences.
In the AdS/CFT correspondence, the conformal anomaly in $N=4$ SYM was seen
on the AdS side as an infrared effect coming from the necessity of truncating
the $AdS_5$ to finite size and choosing a particular conformal structure \witte
\koshen . This very closely parallels our discussion above in the
topological framework.
We note that the central charge
which measures the conformal anomaly in these cases also had
only genus zero and one contributions (of which only the genus zero or
supergravity contribution has been verified).

\subsec{Wilson Loops}

Being a topological gauge theory, the local
observables of Chern-Simons theory are trivial.  The absence of
local degrees of freedom
in the gauge theory is nicely matched with the absence of a physical
graviton
in the dual closed string theory.\foot{However one can write
down a topological gravity theory corresponding to it, known as
``Kahler Gravity'' \ref\bsad{M. Bershadsky and V. Sadov, ``Theory of K\"ahler
Gravity,'' Int. J. Mod. Phys. {\bf A11} (1996) 4689.}.}  Instead, one can
consider Wilson loop observables in the Chern-Simons theory.
In the previous section,
we made the conjecture that these are evaluated on the
closed string theory side by summing over holomorphic maps ending on any loop,
in the boundary $S^3$, which is isotopic to the given loop.
If this is true, then there
is a non-trivial zero'th order check:  if we take the most
general knot invariant on the Chern-Simons side and expand
it in the large $N$ limit, does it admit being written in the form
\dwil , that we expect on the closed string side? This is what we
will verify here, leaving the more detailed checks for later.
Indeed, $<W(\Gamma)>$ for $SU(N)$ give the knot invariants \witten\ studied
by mathematicians. More precisely, it is the HOMFLY polynomial
\ref\matknot{A general reference on knot theory is ``Knots and Physics'',
by L.Kauffman, World Scientific (1993)}
which is traditionally written in terms
of 2 variables, $P_{\Gamma}(q,z)$ with $q=e^{i\pi N\over k+N}$ and
$z=2i\sin({\pi \over k+N})$.
It is interesting that when expressed in
terms of closed string variables, these precisely parametrize the two
independent quantities
$$q={\rm exp}({t\over 2}) ,\qquad z=2i\ {\rm Sin} (g_s/2)$$

For instance, for the simplest loop $C$ which
is topologically trivial -- not knotted at all, the answer is
\eqn\wloop{\eqalign{\langle W(C)\rangle &= {(q-q^{-1})\over z}= {e^{{t\over
2}}-e^{-{t\over
2}}\over 2i{\rm sin}{g_s\over 2}} \cr &= e^{{(t-i\pi)\over 2}}(1-
e^{-t})({1\over g_s}+ \sum_{k=0} c_k g_s^{2k+1}).}}
where the $c_k$'s are some coefficients involving the Bernoulli numbers.
This certainly
looks like the instanton sum in \dwil,
with the leading term coming from the disc (the ${1\over g_s}$ term)
and higher order terms being the analogue of the degenerate higher genus
contributions that we saw in the partition function.
We also see that there are only contributions with $n=0,1$
coming with opposite signs. The overall factor can be thought of as the
framing ambiguity mentioned in Sec.3 .
It would be interesting to understand all these features more precisely.\foot{
Already the answer for the trivial knot given above seems suggestive:
If we can deform the knot at infinity to a circle over $S^2$, the two
contributions at the level of the disc should correspond
to which way
one
wraps around $S^2$ as one bounds the circle.
Moreover a disc
with handles attached would need to have moduli, in order to reproduce
the $Sin$ contribution in the denominator.  In fact this is very much
reminiscent of
the Schwinger-like contribution which gives the contribution of
maps from higher genus to sphere, in that it is roughly speaking,
a square root of it.}

For a general knot, the answer on the gauge theory side follows from the
skein relations shown in figure 2.

\bigskip
\centerline{\epsfxsize 6.truein \epsfysize 1.5truein\epsfbox{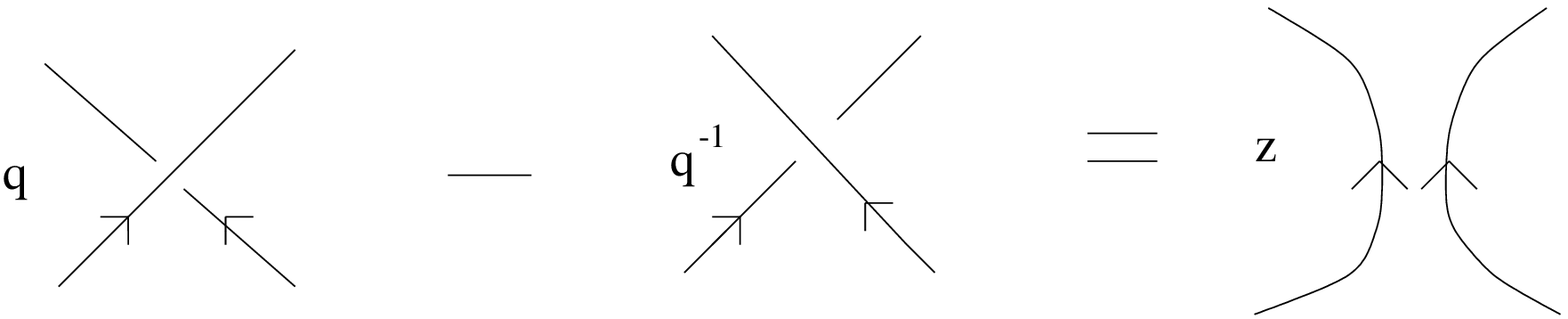}}
\noindent{\ninepoint\sl \baselineskip=8pt {\bf Fig.2}: {\rm
The skein relations }}

The figure depicts manipulations on knots that
can be performed at every crossing. The skein relations
should be viewed as the Chern-Simons analog of the loop equations studied
in connection with Wilson loops.
These relations can be used inductively
to evaluate $<W(\Gamma)>$ in terms of the simple loop average above.
We see that we get extra factors of $z$ (or $g_s$)
when we increase the number
of disjoint components by the operation on the right hand side. Therefore,
the leading term in $W(\Gamma)$ for a single component loop will always be
$O({1\over g_s})$, as expected from a disc. The expansion will also have an
odd parity with respect to $z$, i.e. with only odd powers of $g_s$.
Finally, using the skein relation and \wloop\ it is possible to convince
oneself that only even powers of $q$ appear in $W(\Gamma)$ for a single loop
(upto an overall $q^\alpha$). Noting that $q=e^{t\over
2}$, this has exactly the structure of \dwil\ upto the
overall framing ambiguity.

\newsec{The Linear Sigma Model and Derivation of the Large N/Gravity
Correspondence}

As we reviewed in Sec. 3, for gauge theories arising from D-branes, there are
two limits which are relatively simple. One is the limit when
$\lambda =g_sN\rightarrow 0$ and the other when $\lambda >>0$.  Both
limits are conjectured to be described
by the same underlying string theory with the same small
$g_s$, but with a modified
geometry. Thus, this is not really an S-duality. Rather, it is closer in
spirit
to the different worldsheet descriptions that appear in
different regions of target space parameters, as in Calabi-Yau/Landau-Ginzburg
duality. And one would hope to be able to prove this duality along similar
lines. Having said that, we should point out an important difference.
A closed string sigma model description
when $\lambda >>0$ is somehow going over to an open string one by the time
$\lambda \rightarrow 0$. In other words, holes develop on the originally
closed string world sheet. Is there
an underlying dynamical principle that can give rise to this?
We will now propose a closed worldsheet description, in terms of a linear sigma
model, for all $\lambda$, {\it including
the limit } $\lambda \rightarrow 0$.  What will be special
about the $\lambda \rightarrow 0$ limit is that, the worldsheet
configurations in the IR will consist of
two phases in coexistence.  In one region the 2d
theory becomes
trivial in the infrared and can be integrated out, whereas in the other,
it corresponds to the sigma model on $T^*S^3$ with the $S^3$ shrunk
to a point. Moreover, the fields on the boundary between these two regions
will take values on the $S^3$ in target space.
{\it Thus, in the IR limit we seem to recover both the open string
description with D-branes as well as the closed
string description from a single underlying two dimensional quantum
field theory, simply  by varying a parameter.}

\subsec{The Stringy Description of Chern-Simons on $S^3$}
As reviewed in Sec.2, the Chern-Simons theory on $S^3$ arises from the open
string A-model in the vicinity of the conifold $T^*S^3$.
It is convenient to describe the conifold explicitly as a subspace
in ${\bf C}^4$, defined by the relation
$$x_1x_2-x_3x_4=\mu$$
The space has a Lagrangian $S^3$ submanifold of size $\mu$.
(The $S^3$ can be seen quite explicitly after a change of variables
to $\sum_{i=1}^4 z_i^2=\mu$.
Taking $z_i,\mu$ to be real gives the $S^3$ in the conifold.)
The $\mu$ parameter changes the complex structure of the Calabi-Yau $T^*S^3$.
However, the A-model topological string is independent of
the complex structure of the manifold.
In particular, all amplitudes in the topological
A-model are strictly independent of $\mu$.
We will find it convenient to consider the limit where $\mu\rightarrow 0$.
In this
limit, the equation becomes
$$x_1x_2-x_3x_4=0$$
and the $S^3$ has shrunk to a single point $x_i=0$ for all $i$. Therefore,
in this
limit the topological amplitudes for the A-model will come from open
worldsheets with Dirichlet
conditions $x_i=0$ on the boundary.

\subsec{The Linear Sigma Model Description of a Blown-up Conifold}

There exists a linear sigma model
-- an $N=2$ supersymmetric  $U(1)$ gauge theory -- whose low energy
dynamics description reduces to the usual non-linear
sigma model on the $S^2$ blown up version
of the conifold \wittenl .
One considers an  $N=2$ $U(1)$ gauge theory with four charged
chiral fields, $a_i,b_i$ ($i=1,2$) and charges $+1,-1$ for $a,b$ respectively.
In addition the action has a FI D-term with strength $r$,
as well as a $\theta$ term $i\theta \int F$ for the $U(1)$ gauge field.
As far as the topological theory is concerned, all
quantities will appear in the complex combination $t=r+i\theta$.
Thus having $\theta \not=0$ is equivalently to having
$r\not=0$.
Then,
it was shown in \wittenl\ that this model describes,
in the IR limit, the blown up conifold, where one has an $S^2$
with kahler parameter $t$.  The basic idea is that
there is a D-term potential of the form
$$V_D=e^2(|a_1|^2+|a_2|^2-|b_1|^2-|b_2|^2-r)^2$$
Let us consider $r\not =0$ (say $r>0$), then the
potential forces vacuum configurations
to have either $a_1$ or $a_2$ to be
non-zero. In other words, we are in the Higgs phase. $a_1/a_2$ can be
viewed as the complex coordinate of the $S^2$,
and the two complex normal directions to the $S^2$ can be identified with the
gauge invariant combinations
$b_1a_2$ and $b_2 a_2$.
Thus the target geometry
of the blown up conifold is the same as the Higgs phase of this
gauge theory. (A similar story holds for $r<0$, with the $a$'s and $b$'s
interchanged.)
To see the complex structure of the conifold,
note that if we consider the gauge invariant fields
$$x_1=a_1b_1 ,\qquad x_2=a_2b_2 , \qquad x_3=a_1b_2 , \qquad x_4=a_2b_1$$
the Higgs branch is identified with the gauge invariant
observables, modulo the relation
$$x_1x_2-x_3x_4=0$$
which is the defining equation of the conifold.  The FI D-term
corresponds to the same complex manifold given by the above
equation but with a blown up $S^2$.

The $N=2$ gauge theory has in addition, a
complex neutral scalar field $\sigma$ in the $U(1)$ gauge multiplet.
In the Higgs branch this $\sigma$
field is massive, due to interactions of the form
$$V_{int}= e^2|\sigma |^2(|a_1|^2+|a_2|^2+|b_1|^2+|b_2|^2)$$
When $r\not =0$, the Coulomb branch,
corresponding to $<\sigma >\not=0$ is absent, because
the FI term necessarily Higgses the $U(1)$ and gives
masses to the $\sigma$ field due to the above interactions.
Thus the $\sigma$ field is irrelevant in the IR limit.

Now, let us consider the limit where $t\rightarrow 0$. In this limit,
the low energy configurations allow both the Higgs and the Coulomb
branch to coexist. In the Higgs branch we have the description of
the conifold as the moduli of vacuum configurations, but now with
no $S^2$ blown up.  In other words it is at the point of the conifold
transition.
Just as before, the $\sigma$ field is massive in the Higgs branch.
However, we now also have the Coulomb branch, in which case
$<\sigma >\not= 0$ and where all the charged fields $a_i,b_i$ are massive
and frozen in the infrared limit at $a_i=b_i=0$.  The Coulomb
branch is in fact a free $N=2$ abelian $U(1)$ gauge theory and
therefore a trivial free theory in the infrared.

\subsec{The Topological Sigma Model and the Emergence of D-branes}

The issue now is to dynamically see the emergence of worldsheets with
holes (and Dirichlet boundary conditions) as
$t\rightarrow 0$.
As we have
seen from our discussion of the linear sigma model describing the
conifold, it is exactly in this limit that
the system will have two branches:
the Higgs branch and the Coulomb branch.
In $d>2$ one simply fixes the vevs of fields so that one is
on one branch or
the other. As is well known, the situation
in $d=2$ is different, and we have to allow all possible fluctuations.
In fact, in our case, as $t\rightarrow 0$,
where we have both a Higgs (H) and a Coulomb (C) branch,
we can write down spatial configurations which at $x=-\infty$ start
on one branch and at $x=\infty$ end on the other.  Moreover, one
can arrange the energy of this configuration to be as small as one
pleases.\foot{
We would like to thank Sidney Coleman for a discussion on this point.}.
The basic point is that if we are in one branch, we can
change vevs slowly enough over large regions so as
to cost little energy.
In this way, both phases coexist in the worldsheet fluctuations
even in the IR.
Note, in particular that the region for each phase
could be as big as one wishes it to be.  Thus at $t=0$ we have to
consider worldsheets consisting of large regions of different phases (H)
and (C).  Approaching $t=0$ from positive values, we would have started
by only having the (H) branch, and as $t$ become smaller we can get
arbitrary fluctuations involving patches in the (C) branch.
As we noted earlier,
all the (charged) fields $a_i,b_i$
pick up a mass in the (C)
branch and are irrelevant in the IR. In fact, their values are frozen to zero.
Actually, the (C) branch in the IR is
a free abelian theory
which can be integrated out leaving
us with an effective theory in the (H) branch with gaping
holes corresponding to
the (C) patches. Moreover,
the fact that the charged fields vanish in the (C) branch
gives Dirichlet boundary condition to the fields $a_i=0$ and $b_i=0$
at the boundary. In other words, we have
the same D-brane boundary conditions that we would have
expected on the Chern-Simons side in the limit
when the $S^3\rightarrow 0$.
We are thus seeing the D-branes
anticipated from our duality emerging from the worldsheet description.

For this picture to be the right one,
we also need to show how the open string factor of $(g_s N)^B$
multiplying worldsheets with $B$ boundaries arises on the
closed string theory side, in the limit where $g_sN\rightarrow 0$.
Note that we are identifying $B$ with the number of (C) patches of the
linear
sigma model.
We will offer a speculative explanation of how this might arise.
It will
be important to verify this idea in a more detailed study.

Recall the identification,
$g_sN=\l=Im t=\theta $ in the $U(1)$ gauge theory
($\theta$ appears in the action only in the term $\theta \int F$).
Let us assume that the (C)
patches can
be in topological sectors with vortex charge of either zero or one.
In other words, since we do not have multiply charged bound vortices
we exclude multiple vortex field
configurations in a given Coulomb patch. If we now do a path integral
in this free theory, the boson and fermion determinants will cancel up to a
sign.
Let us assume that in the one vortex sector we
get an extra minus sign from determinants, compared to the one where
the vortex field is absent.  Thus the contribution from each
Coulomb patch will be
$$f=1-{\rm exp}(i\theta)$$
In the limit $\theta \rightarrow 0$, we have $f\sim \theta$ and so
we get a factor of $f^B=\theta^B=(g_s N)^B$ from integrating
out all the $B$ Coulomb patches.  Note that the factor $1-{\rm exp}(i\theta)$
is the quantum deformation of $\theta$, in the sense of quantum groups
where
$$q-q^{-1}\rightarrow i\theta$$
with $q=exp(i\theta/2)$.  Note also that the contribution
$1-exp(i\theta)$ is suggestive of the fact, discussed
in Sec.3  that a simple Wilson loop
is expected to give $1-q^2$ (up to an overall factor of
$q^\alpha$).

\newsec{Conclusions and Generalizations}

We have seen encouraging signs that the large N/Gravity correspondence
involving D-branes might actually be provable at the level of string
perturbation theory, i.e. in terms of
certain facts about 2d QFT's.  There are clearly aspects
of our argument in our particular
example which need to be further developed. In particular,
we obtained Dirichlet conditions from the closed string side
in the $S^3\rightarrow 0$ limit. But one would like to check
that the deformation away from this degenerate limit does not destroy
the correspondence. Also it is important to verify our
picture of how $g_sN$ factors arise from each hole.

We have been fortuitous to get an explicit closed string description
of a large $N$ gauge theory which admits completely independent checks
on both sides. We expect that even this case, simple as it may be, will
give us more insights into the general nature of the gauge theory/geometry
correspondence. It is heartening that supersymmetry did not play any
crucial role in our considerations.
There are also other aspects of this Chern-Simons/topological closed string
correspondence that deserve study such as the detailed understanding of
Wilson Loops. This might also be a mathematically fruitful connection
to make.  Also it would be interesting to elucidate the role
of the finite size $S^2$ in the gauge theory.

Another natural question to consider is whether the closed
topological string on a
compact
Calabi-Yau can be dual to a gauge theory.  If there is such
a description, given the IR/UV relation,
and given that the large distance is naturally cutoff in the compact
Calabi-Yau, one would expect the gauge theory to have a natural
short distance cutoff.  It would be interesting to explore
such a possibility in connection with integrable lattice models
and their relation to Chern-Simons theory.

 There are various immediate generalizations of what we considered
in this paper.  For example, we can consider Chern-Simons
theory on Lens spaces (or
their generalizations corresponding to quotients of $S^3$ by discrete
subgroups of $SU(2)_L\times SU(2)_R$ acting on $S^3$) and develop
closed string duals (some aspects of these theories were discussed in
\gopvaf ).
One could also consider large $N$ $SO$ and $Sp$ gauge groups in a similarly
explicit manner.
The topological B-model in this closed string geometry is also worthy of study.
(For example, by wrapping $D_2$ branes on the $S^2$ and considering closed
strings on the $T^*S^3$ geometry, we would be studying the
mirror of our present example).
Another interesting and plausibly tractable case is that with the
$N=4$ topological
string \ref\berkv{N. Berkovits and C. Vafa, ``N=4
Topological Strings,'' Nucl. Phys. {\bf B433} (1995) 123.} which should be
related to 2d field theories
(and in particular to Large $N$ principal chiral models which have
an interesting and exactly solvable mass spectrum etc.)
The closed string  dual in this case describes
self-dual gravity
theories in 4d which is also the $N=2$ string theory
\ref\oov{H. Ooguri and C. Vafa, ``Geometry of N=2 Strings,'',
Nucl. Phys. {\bf B361} (1991) 469.}. This
raises the fascinating possibility of understanding the $(2,2)$ signature
self-dual
gravity theory in the large $N$ limit of a $(1,1)$ signature QFT.
It might also be of interest to re-examine the
string description of $QCD_2$ \moor\
and perhaps make it as explicit as the gauge theory side (See the recent
paper of Horava in \moor\ ).

Finally, we believe the linear sigma model approach might help
in deriving the
AdS/CFT correspondence.  Namely, by considering  a linear
sigma
model for an AdS background
with a RR flux turned on, we expect to automatically end up with the
D-brane rules of open string theory in the limit where $g_sN\rightarrow 0$,
presumably via a two phase system.
This would then be a proof of the AdS/CFT correspondence
using perturbative string theory techniques.
The main task is to first find a convenient linear
sigma model description of the AdS background with a RR flux. Then it is
a matter of dynamics to determine whether or not the behaviour as
$g_sN\rightarrow 0$ allows us to, say,  integrate a trivial phase out,
leaving us effectively with a worldsheet with Dirichlet boundaries.
Moreover, the fact that we need
to take $\alpha'\rightarrow 0$ is somewhat similar to having
a topological limit on the gauge theory side.

We would like to thank M. Bershadsky, S. Coleman,  S. Katz, A. Lawrence,
J. Maldacena, T. Pantev,
A. Strominger, C. Taubes and S.T. Yau for valuable discussions.

The research of R.G. is supported by DOE grant
DE-FG02-91 ER40654 and that of C.V. is supported in part by NSF grant
PHY-98-02709.

\listrefs

\end